\documentclass{mn2e}
\usepackage{graphicx}
\usepackage{epsfig}
\usepackage{epsf}
\usepackage{deluxetable}

\title[SN 1998S 14 years Postmortem]{Supernova 1998S at 14 years Postmortem: Continuing Circumstellar Interaction and Dust Formation}
\author[Jon Mauerhan and Nathan Smith]{Jon Mauerhan$^{1}$\thanks{E-mail:
mauerhan@as.arizona.edu; nathans@as.arizona.edu} and Nathan
Smith$^{1}$\footnotemark[1]\\
$^{1}$University of Arizona, Steward Observatory, Tucson, Arizona  85721, USA}
\begin{document}

\date{Submitted to MRAS April 6 2012.}

\pagerange{\pageref{firstpage}--\pageref{lastpage}} \pubyear{2012}

\maketitle

\label{firstpage}

\begin{abstract}

  We report late-time spectroscopic observations of the Type~IIn supernova (SN) 1998S, 
  taken 14 years after explosion using the Large Binocular Telescope. The optical spectrum 
  exhibits strong broad emission features of [O {\sc i}], [O {\sc ii}], and H$\alpha$, in addition to 
  weaker features of [O {\sc iii}], H$\beta$ and [Fe {\sc ii}].  The last decade of evolution has exhibited 
  a strengthening of the oxygen transitions relative to H$\alpha$, evidence that the late-time emission 
  is powered by increasingly metal-rich SN ejecta crossing the reverse shock. The H$\alpha$ luminosity 
  of $\approx$8000 $L_{\odot}$ requires that SN 1998S is still interacting with relatively dense circumstellar 
  material (CSM), probably produced by the strong wind of a red supergiant progenitor at least $\sim10^3$ 
  years before explosion.  The emission lines exhibit asymmetric blueshifted profiles, which implies that the 
  receding hemisphere of the SN is obscured by dust. The [O~{\sc iii}] $\lambda$5007 line, in particular, 
  exhibits a complete suppression of its red wing. This could be the result of the expected wavelength dependence 
  for dust extinction or a smaller radial distribution for [O~{\sc iii}]. In the latter case, the red wing of [O~{\sc iii}] could be absorbed by core dust, while both the blue and red wings are absorbed by dust within the cool dense shell between the forward and reverse shocks; this interpretation could explain why late-time [O~{\sc iii}] emission from SNe is often weaker than models predict. The [O {\sc i}] line exhibits  double-peaked structure on top of the broader underlying profile, possibly due to emission from individual clumps of ejecta or ring-like structures of metal-rich debris. The centroids of the peaks are blueshifted and lack a red counterpart. However, an archival spectrum obtained on day 1093 exhibits a third, redshifted peak, which we suspect has become extinguished by dust that formed over the last decade, after day 1093. This implies that the ``missing" red components of multi-peaked oxygen profiles observed in other SNe might be obscured by varying degrees of dust extinction. 
 \end{abstract}


\section{Introduction}

Supernova (SN) 1998S, which occurred in the nearby host galaxy
NGC~3877, is one of the most well-studied SNe of the core-collapse variety. It
is designated a Type~IIn, a class which has raised critical new
questions about the final phases of massive-star evolution.  The
``IIn'' designation refers to narrow emission lines of hydrogen
observed in optical spectra (see Fillipenko 1997), caused by
relatively slow-moving circumstellar gas that is impacted by the blast
wave or illuminated by UV radiation from the SN.  The presence of
these narrow lines requires dense circumstellar material (CSM) ejected
by the progenitor star shortly (within a few years to decades) before
the explosion.  In some extreme cases, up to 25 M$_{\odot}$ has been
ejected in the decade before core collapse (Smith et al.\ 2007, 2008,
2010; Woosley et al.\ 2007), which has suggested a possible link to
the class of eruptive luminous blue variables (LBVs). The impact of
the SN blast wave on the CSM can efficiently convert kinetic energy
into radiation, which explains why some Type IIn events have been associated with
several of the most luminous SNe yet recognized in the Universe, such as SN~2006gy (Smith et al. 2007, 2010; Ofek et al. 2007).
In other less extreme cases, such as SN 1998S, SNe IIn can exhibit interaction between 
the blast wave and a dense, but relatively-steady, hydrogen-rich wind from 
a red or blue supergiant progenitor.   In all cases, long-term monitoring of the spectral evolution
of Type~IIn SNe present the opportunity to probe the geometry,
kinematics, structure, and chemistry of both the slow-moving CSM and the rapidly
expanding SN ejecta.

Early spectra of SN~1998S showed signs of CSM interaction very soon after explosion.   Typical of Type~IIn evolution at early
times, relatively narrow H$\alpha$ and He {\sc i} emission lines
exhibited broad Lorentzian wings, 
probably formed by electron scattering of photons diffusing out of the opaque 
pre-shock gas (Leonard et al. 2000; Chugai 2001; Fassia et al. 2001). 
Narrow P-Cygni profiles developed in the UV and optical spectra, generated 
by the slow-moving ionized stellar wind of the progenitor, 
which had an estimated velocity of 40--50 km s$^{-1}$ (Bowen et al.\ 2000; Fassia et al. 2001). At
about 200 days, the H$\alpha$, H$\beta$, and He {\sc i} lines began to
exhibit complex structure in the form of asymmetric multi-peaked
profiles, interpreted as the result of an expanding disk-like
structure within the outflow (Leonard et al.\ 2000; Gerardy et al.\
2000). SN~1998S also showed signs of thermal dust emission in the IR,
either produced by the SN itself, thermal reprocessing
of SN radiation by CSM dust, light echoes, or some combination of these (Fassia et al. 2000; Gerardy et
al. 2002). The SN was also a strong long-lasting source of thermal
X-rays, and generated radio free-free emission after about 300 days
(Pooley et al. 2002), additional signposts of a lively CSM
interaction. 

An important question concerning SNe~IIn is how closely the eruptive
pre-SN mass loss is linked to the impending core-collapse itself, or
for how long before explosion the progenitor exhibited an unusually
high mass-loss rate.  For Type~IIn SNe, as well as for all other SNe
that may explode into dense CSM, the impact of the blast wave upon the
CSM can potentially manifest itself in observable characteristics that
persist for decades. Such SNe, not necessarily classified specifically as Type IIn,  include 1957D, 1970G, 1979C, 1980K, 1986E, 1986J, 1993J, 1996cr, 1988Z, and of course SN 1987A (see Milisavljevic et al. 2012, and references therein).
 Here we present data obtained 14-years after explosion, which adds SN~1998S to this group of persistently interacting core-collapse SNe, and we discuss this object's emission-line features in context.

\begin{figure}
\includegraphics[width=3.3in]{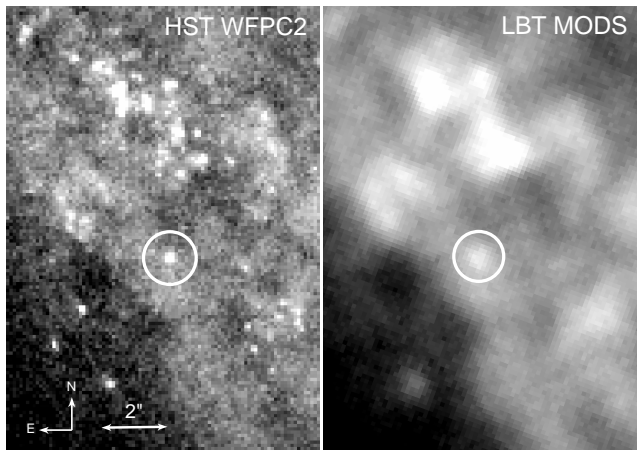}
\caption{The local field of SN~1998S. The \textit{HST}/WFPC2
  image (F814W, \textit{left}) is from 2000 Oct 1, and the LBT/MODS
  image ($g$-band; \textit{right}) was obtained 2012 Jan 27.}
\label{fig:sn_im}
\end{figure}

\section{Observations}

SN 1998S was targeted for spectroscopic observation on 2012 Jan 27
(day 5079 after discovery) with the Multi-Object Double Spectrograph
(MODS; Byard \& O'Brien 2000) on the 8 m Large Binocular Telescope
(LBT). Under $\sim$0\farcs8 seeing, a source at the SN position was
detected in the 60.0 sec $r$-band target acquisition image, shown in
Figure~\ref{fig:sn_im}.  From this image we were able to extract a photometric
measurement using aperture photometry, deriving an $r$-band (SDSS)
brightness of 23.3$\pm$0.3 mag.

Spectral images in the blue and red channels were obtained with MODS
in longslit mode, utilizing the G670L grating and a 1\farcs2 slit,
which yielded a spectral resolution of $R\approx$1000. Because of
guide star constraints, we had to observe the SN with the slit at
$\approx$45$^{\circ}$ from the parallactic angle, prompting the concern 
that some blue light could have been lost as a result. However, since the SN
was observed at an airmass of only $\approx$1.1, and a 1\farcs2 slit
was used under $\approx$0\farcs8 seeing, we do not expect substantial 
loss of blue flux.  Four exposures totaling 80 min were obtained at the SN position. 

The spectral images
were bias subtracted, flat fielded, and median combined with a
suitable rejection filter to remove cosmic rays.  The spectrum was
extracted using standard {\sc{IRAF}} routines. Removal of the background line
emission from the host galaxy required us to sample the background 
from a narrow strip very close to the dispersion track of the SN
emission, such that the background region overlapped very slightly
with the wings of the SN point spread function (PSF), but the result 
subtracted less than a few percent of the SN flux. The extracted spectrum 
was wavelength calibrated using spectra of HeNeAr lamps. The wavelength 
solution was corrected for the redshift of NGC 3877 ($z=0.003$, Tully \& Fisher 1988).

\section{Results}
\begin{figure}
\includegraphics[width=3.3in]{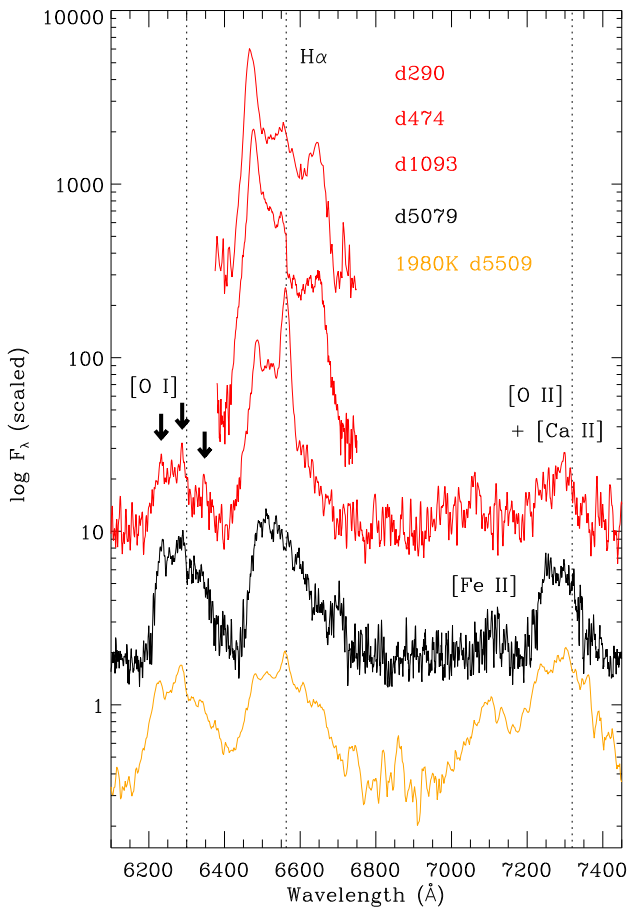}
\caption{The LBT/MODS red channel spectrum of SN~1998S obtained 2012 Jan 27
  (day 5079, \textit{black}), compared with spectra of the same source
  on days 290, 474, and 1093 (\textit{red}; Pozzo et al. 2004; Fassia et al. 2001).  
  A spectrum of SN 1980K on day 5509 is also included for comparison (Fesen et al. 1999).
  On day 1093, the [O {\sc i}] profile exhibited a triple-peaked
  narrow-component (\textit{black arrows}), but the reddest of these has diminished by day 5079.}
\label{fig:spec_comp}
\end{figure}

\begin{figure}
\includegraphics[width=3.3in]{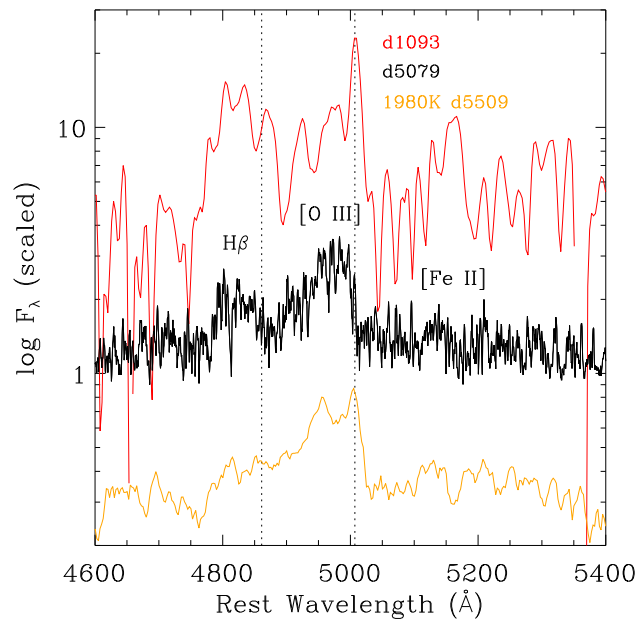}
\caption{LBT/MODS blue channel spectrum of SN 1998S on day 5079 (\textit{black}), compared to day 1093 (\textit{red}, Pozzo et al. 2004), and SN 1980K 5509 days after discovery (\textit{orange}, Fesen et al. 1999) .}
\label{fig:spec_compb}
\end{figure}

The new LBT/MODS spectrum of SN~1998S from day 5079 is shown in Figure~\ref{fig:spec_comp}, 
where it is compared to three earlier epochs from day 290, 474, and 1093 
(archival spectra from Pozzo et al. 2004, obtained from the Weizmann interactive supernova data repository\footnote{www.weizmann.ac.il/astrophysics/wiserep}; Yaron \& Gal-Yam 2012).
For days 290 and 474, only the range of the H$\alpha$ line is presented, since there was a lack of significant detections from other
transitions at these epochs. Figure~\ref{fig:spec_compb} shows our blue channel spectrum of SN 1998S and the earlier day 1093 epoch. For comparison,  we included spectra of SN~1980K at 5509 days after discovery (Fesen et al. 1999; spectrum kindly provided by R. Fesen and D. Milisavljevic).
Although the MODS wavelength coverage extends beyond the regions
displayed in Figure 2, barely any continuum emission was detected on
the latest epoch of SN~1998S outside of the displayed wavelength
range. This insignificant amount of continuum is likely contaminated with
 background emission from the H {\sc ii} region that SN 1998S originated in,
 so we refrain from interpreting it.

The red channel spectra are dominated by strong, broad emission lines
of H$\alpha$, and the forbidden doublet transitions of
[O {\sc i}] $\lambda\lambda$6300, 6364 and [O {\sc ii}]
$\lambda\lambda$7319,7330. There is also a
possible contribution from [Ca {\sc ii}] $\lambda\lambda$7291, 7324
which overlaps with [O {\sc ii}], although the latter is likely to be the dominant transition (Fesen et al. 1999). 
A weak but broad feature of [Fe {\sc  ii}] $\lambda$7155 is also detected on the blue side of [O {\sc ii}],
although [Ar {\sc iii}] $\lambda$7136 might be contributing to the flux as well.

The blue channel is dominated by [O {\sc iii}] $\lambda$5007 and
H$\beta$ $\lambda$4861, with the former exhibiting a dramatic cutoff of its blue wing.
 Weak [Fe {\sc ii}] emission is also present around $\lambda$5100--5200, 
 where multiple transitions of this species lie. 

Emission-line intensities, luminosities, and velocity parameters are listed in
Table~1.  The full-width at zero intensity  (FWZI) of the H$\alpha$ line is $\approx$240 {\AA}, or $\approx$11,000 km
s$^{-1}$, while the oxygen lines have FWZI values closer to $\approx$8000--9000 km
s$^{-1}$. For H$\beta$ and [O {\sc iii}], the full extent of their respective red and blue wings is difficult to ascertain, 
because of the partial overlap of these lines. Luminosity calculations were made after applying foreground extinction and distance estimates of $A_V=0.68_{-0.25}^{+0.34}$ mag (Fassia et al. 2001) and $15.5\pm1.3$ Mpc (NASA/IPAC Extragalactic Database [NED]), respectively. Extinction values were calculated for the rest wavelength of the emission line in question, assuming an extinction coefficient of $R$=3.1 (Cardelli, Clayton, \& Mathis 1989).  No correction was made for the uncertain degree of internal 
extinction.  \medskip \\

All of the line profiles exhibit asymmetric blueshifted peaks, probably the result of a suppression
of flux on their red wings. The flux drop is 
particularly steep for [O {\sc iii}] and also for the central portion of the [O {\sc i}]
profile, with flux diminishing abruptly longward of their rest wavelengths. 
\begin{center}
\begin{table*}
{\small
\hfill{}
\caption{Line Measurements for SN 1998S on day 5079 ($\approx$14 yr). Luminosity errors are $\approx$0.15 dex, mainly dependent on distance and extinction uncertainties (day 2148 error is 0.25 dex, see text). Line ratios for the spectroscopically similar SNe 1980K and 1979C, at ages of 14--15 yr and 12--14 yr, respectively, are included for comparison (Fesen et al. 1999). Approximate luminosity ratios are given with respect to H$\alpha$. CF94 model predictions are included in the last column for power-law, and lower limits for RSG models.}
\begin{tabular}{@{}lcccccccc@{}}
\hline
   Line    &  $\lambda_{0}$ &  Intensity & $\Delta V$ (edges)  & log $L$ & \multicolumn{4}{c}{$L/L$(H$\alpha$)}  \\                            
               &  ({\AA})         &   (erg cm$^{-2}$ s$^{-1}$) & {(km s$^{-1}$)} & (erg s$^{-1}$) & (SN 1998S) & (SN 1980K) & (SN 1979C) & (CF94 model)\\
\hline
H$\beta$     & 4681      & $6.3\times10^{-17}$      & \nodata & 36.6 & 0.1 &  \nodata & \nodata & 0.3\\
$[$O {\sc iii}$]$ & 4959, 5007      & $1.3\times10^{-16}$      & $-5500$, $<$100 & 36.9  & 0.3 & 0.1--0.2 &  0.9--1.1 & 3.4, $>$3.4 \\
$[$O {\sc i}$]$   & 6300 6364  & $3.7\times10^{-16}$    &  $-4500, 4000$ & 37.3 & 0.6 & 0.7 & 0.9 & 0.6, $>$0.1 \\
H$\alpha$         & 6563      & $6.4\times10^{-16}$     & $-5500, 6000$$+$& 37.5  & 1 & 1 & 1 & 1 \\
$[$Fe {\sc ii}$]$ & 7155      & $3.5\times10^{-17}$      & \nodata & 36.2 & 0.05 & \nodata & \nodata & \nodata \\
$[$O {\sc ii}$]$$+$$[$Ca {\sc ii}$]$ & 7319, 7330      & $2.6\times10^{-16}$     & $-5500, 3500$&  37.1 & 0.4 & 0.9--1.0 & 11--12 & 0.4, $>$3.5 \\
\hline
\end{tabular}}
\hfill{}
\label{tb:tablename}
\end{table*}
\end{center}

There are conspicuous changes between day 1093 and 5079. Perhaps the most obvious are the disappearance of the
intermediate-width, systemic-velocity components of the
H$\alpha$ line, and the increasing strength of oxygen
emission with respect to H$\alpha$. Less obvious is the change in the triple-peaked profile
of the [O {\sc i}] $\lambda\lambda$6300, 6364 emission feature. The central peak, assuming it
traces the $\lambda$6300 component of the doublet, appears
blueshifted from rest by roughly $-$610 km s$^{-1}$, and the blue and
red peaks are shifted by $\approx-2600$ and $+2800$ km s$^{-1}$, respectively, from the central peak. It seems likely, although not  certain, that each of
these peaks represent different velocity components of the
$\lambda$6300 transition of the doublet, since the underlying broad emission is centered near $\lambda$6300, with no comparable broad emission centered near $\lambda6364$. However, if the red peak is instead from the $\lambda$6364 component of the doublet, then it is exhibiting a blueshift of $-$800 km s$^{-1}$.  Although the change is
subtle, the central peak appears to become more sharply cutoff on its
red side by day 5079. The blueshifted peak remains more or less
unchanged. Remarkably, the reddest peak is no longer detected on day 5079.

\begin{figure}
\includegraphics[width=3.4in]{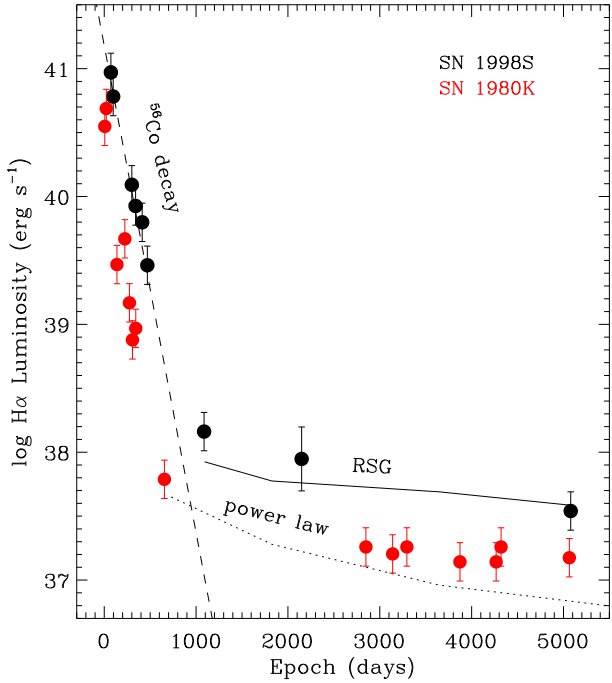}
\caption{H$\alpha$ luminosity curves for
  SN~1998S and SN~1980K. The luminosity at epochs $<$1000 days
  is consistent with the 111.5-day mean lifetime of radioactive $^{56}$Co 
 (\textit{dashed line}), but later epochs from day 1093 and onward fall above this trend by as much as 10 dex,
requiring CSM interaction.  The late-time luminosity of SN 1998S 
appears most consistent with model predictions for an RSG density profile (CF94; \textit{solid line}), while
 SN 1980K lies between this and a power-law density profile (\textit{dotted line}), as discussed in \S4.}
\label{fig:lc}
\end{figure}

The most recent epoch on day 5079 exhibits $L$(H$\alpha$)=$(3.1\pm1.4)\times10^{37}$ erg s$^{-1}$ ($\approx$8000 $L_{\odot}$). To compare this with past luminosity values, we adopted flux measurements from 
Fassia et al. (2001) and Pozzo et al. (2004) from day 1093 and earlier, and converted them to luminosities using the same distance and extinction values as before.  In addition, an H$\alpha$ flux for day 2148 was extracted from a spectrum presented in Fransson et al. (2005; kindly provided by T. Matheson, but not shown here). The day 2148 spectrum was observed under poor seeing conditions. Thus, there was substantial and inseparable contamination from the background H {\sc ii} region, specifically, narrow H$\alpha$ and [N {\sc ii}]-doublet emission. We interpolated over these narrow features in order to estimate the broad underlying H$\alpha$ flux of the SN. As a result, we may have removed some fraction of narrow H$\alpha$ flux potentially intrinsic to the SN, if any were present at rest velocity on day 2148. In light of these caveats, we estimate a conservative error of 0.25 dex for the luminosity on day 2148. For additional comparison, we also include the H$\alpha$ light curve for SN 1980K in Figure~\ref{fig:lc} (See Milisavljevic et al. 2012, and references therein), since this SN has exhibited spectral evolution very similar to SN~1998S, and has excellent late-time coverage. For SN 1980K, we adopted $E(B-V)=0.41$ mag (Fesen et al. 1999) and a distance of 6 Mpc (NED).

Earlier than 1000 days, the light curve for H$\alpha$ is well matched by the 
radioactive decay trend of $^{56}$Co (111.5-day mean lifetime; Firestone et al. 1996). After 1000 days, the light curve departs from
this trend, and hits a floor of $\sim10^{37.5} L_{\odot}$ that is maintained through day 5079.  Since the day 5079 spectrum differs from the spectrum at early times, the late time H$\alpha$ luminosity is not likely to be from an echo. Rather, the luminosity is accounted for by SN/CSM interaction, which we discuss in the following section.

\section{Discussion}  

\subsection{Source of the Broad Line Emission}

The collision between the SN explosion and the CSM can have
lasting effects on the optical emission-line spectrum, long after the
continuum has faded away. The impact of the SN blast wave on the CSM
leads to the formation of a forward shock that meets the progenitor wind, and a reverse shock that
gets bombarded by the expanding metal-rich SN ejecta (Chugai 1991; Chevalier \& Fransson 1994,
hereafter CF94). Gas in the cool dense shell (CDS) that develops between the forward and reverse shocks
 absorbs far-UV and X-ray photons generated within the shock, and this excites low-ionization 
 transitions such as [Fe {\sc ii}] and [O {\sc i}]. Inside the reverse shock, the hot outer layers of the 
 metal-rich ejecta are also ionized, but give rise to higher ionization transitions, such as [O {\sc iii}] 
 and UV metal lines. Both the CDS and the ionized ejecta regions produce comparable amounts of H$\alpha$, H$\beta$, and [O {\sc i}] emission, while [O {\sc iii}]
 is generated almost exclusively on the skin of the ejecta, according to CF94. At late times ($t>10$ yr), the luminosity of the reverse shock becomes significantly higher than the 
 forward-shocked CSM gas (by a factor of $\approx$30, CF94).  As the ejecta expands, [O {\sc i}], [O {\sc ii}],
and [O {\sc iii}] emission are expected to increase relative to H$\alpha$, as collisional de-excitation 
becomes less effective for these forbidden transitions as the density drops. 
 
 CF94 provide predictions for the UV and optical emission-line luminosity that results from SN/CSM interaction. 
 The luminosity of a given line depends on the ionizing luminosity of the reverse shock, which is sensitive to the
  density profile of the expanding supernova gas. This profile is, in turn, dependent on the stellar structure, the wind of the progenitor, and on the shock kinematics. CF94 consider the density profile of a red supergiant (RSG), which has relatively steep density gradients, in addition to more smooth power-law forms. The luminosity dependence of H$\alpha$ can thus be expressed as
  \begin{equation}
L(\textrm{H}\alpha)\propto \frac{L_s}{T_s}\propto \frac{V_s^3}{T_s}\left ( \frac{\dot{M}}{v_w} \right),
 \end{equation} 
 where $L_s$, $T_s$, and V$_s$ are the shock ionizing luminosity, temperature, and velocity, respectively. $\dot{M}/v_w$ is the wind density parameter, 
 where the mass-loss rate and progenitor wind velocity are given in units of $10^{-5} M_{\odot}$ yr$^{-1}$ and 
 10 km s$^{-1}$, respectively.
 
 The H$\alpha$ light curve for SN 1998S is shown in Figure \ref{fig:lc}, and includes model tracks for the SN density profile of a RSG and power-law distribution. The power law curve connects epochs 2, 5, 10, and 17 yr, adopted directly from CF94 (their Table 6); and the RSG curve was generated for epochs 3, 5, 10, and 14 yr by scaling the 10-yr H$\alpha$ luminosity given in CF94
by the time-variable $L_s/T_s$ ratio.  Both models have $\dot{M}/v_w=5$, which is an appropriate wind density parameter for SN 1998S, since the mass-loss rate and wind velocity of the progenitor have been estimated at 40--50 km s$^{-1}$ and 1--2$\times10^{-4} M_{\odot}$ yr$^{-1}$ (Fassia et al. 2001; Pooley et al. 2002). The RSG and power-law CF94 models have ejecta velocities of $4318$ and 5038 km s$^{-1}$, respectively. These velocities imply that 5079 days after explosion the shock should be located $\sim2\times10^{12}$ km from the SN origin ($\sim$13000 AU). Red supergiant winds extend to at least $\sim3\times10^{13}$ km radius (Lloyd, O'Brien \& Kahn 1995). Thus, the ongoing SN/CSM interaction during the most recent epoch is to be expected.  

Overall, the late-time H$\alpha$ luminosity trend of SN 1998S illustrated in Figure~\ref{fig:lc} is apparently most consistent with the RSG model.  The H$\alpha$ luminosity of SN 1980K, on the other hand, has late-time luminosity values $\approx$6 times lower than SN 1998S at similar epochs. The lower values could be explained as the result of the lower mass-loss rate of the SN 1980K progenitor ($2.5\times10^{-5} M_{\odot}$ yr$^{-1}$, Fesen et al. 1999), which is a factor of $\approx$6 lower than SN 1998S. As a result, the wind density parameter and, hence, the reverse shock luminosity for this object would also be lower by a comparable factor. Alternatively, the power-law density profile may be more consistent with SN 1980K, as argued by Fesen et al. (1999). However, although the RSG model is favored in the case of SN 1998S, it would premature to overly discriminate between either models using line luminosities alone, since one only has to tweak the wind density parameter or shock velocity to obtain a reasonable match to the data. It is also important to keep in mind that the luminosity could be affected by the highly uncertain degree of local extinction and/or asymmetries in emitting volume. If the ejecta or progenitor wind are asymmetric, the models we have considered here would be providing an underestimate of $\dot{M}/v_w$ and $L$(H$\alpha$). In any case, the implied mass loss rate of $\sim10^{-4} M_{\odot}$ yr$^{-1}$ for SN 1998S is extremely high, at the far upper end of the range exhibited by RSGs (Gehrz \& Woolf 1971; Jura \& Kleinmann 1990). 

Elevated mass-loss rates could be a sign of impending core collapse for RSGs, perhaps driven by pulsation instabilities that generate a superwind (Yoon \& Cantiello 2010). A suitable Galactic analog to the progenitor of SN 1998S could be the extraordinary object VY CMa (see Smith et al. 2001 and references therein). This RSG star is experiencing a very large mass-loss rate of $\dot{M}=2$--$4\times10^{-4}$ (with possible brief bouts as high as $\sim10^{-3} M_{\odot}$ yr$^{-1}$). It exhibits an average wind speed of 30--40 km s$^{-1}$ and is surrounded by dense CSM that extends  $\sim$1000 AU from the star, estimated to have been ejected over the last $\sim$1000 yrs. For these reasons, VY CMa has  been highlighted as a viable SN IIn progenitor (Smith, Hinkle, \& Ryde 2009a).  For the progenitor of SN 1998S, the wind speed of 40--50 km s$^{-1}$ combined with our estimated shock distance of 13000 AU implies that its high rate of mass loss began at least $\approx$1200 years before explosion. If this is close to the onset time for RSG superwinds (as suggested by Yoon \& Cantiello 2010), and if the entering of this phase signifies a countdown to core collapse, it thus implies that VY CMa (whose superwind has been blowing for $\sim$1000 years) could explode relatively soon.

We observe an increase in oxygen emission relative to H$\alpha$ over the last decade, providing evidence that the late-time emission traces increasingly metal-rich SN ejecta crossing the reverse shock (CF94). Between days 1093 and 5079 the [O~{\sc i}]/H$\alpha$ ratio has increased from 0.1 to 0.6. This is roughly consistent with the power-law model and, to a lesser extent, the RSG model. However, over the same time period the [O~{\sc iii}]/H$\alpha$ ratio increased by a much smaller amount, from 0.2 to 0.3, which is an order of magnitude below the value predicted for the power-law model, and a factor of at least $\approx$17 below the RSG model. Serious discrepancies between the observed late-time line ratios and the CF94 model predictions, particularly regarding [O~{\sc iii}], have been reported for several other SNe observed at late times, including SN 1980K and 1979C (CF94; Fesen et al. 1999).  The ratio is highly sensitive to the exact value of the shock velocity and the ejecta density. CF94 have suggested that a flatter density profile for the high-velocity ejecta, in addition to variations in shock temperature, could result in a lower ratio that would help resolve the discrepancy. Alternatively, the structure of the [O~{\sc iii}] line profiles in SN 1998S and 1980K could be an indication that the optical depth of the [O~{\sc iii}] emitting region beneath the reverse shock is larger than expected, which would also result in a lower observed line ratio; and we explore this possibility further below. The geometry and filling factor of the [O~{\sc iii}] emitting region(s) could affect the result as well.

\begin{figure}
\includegraphics[width=3.4in]{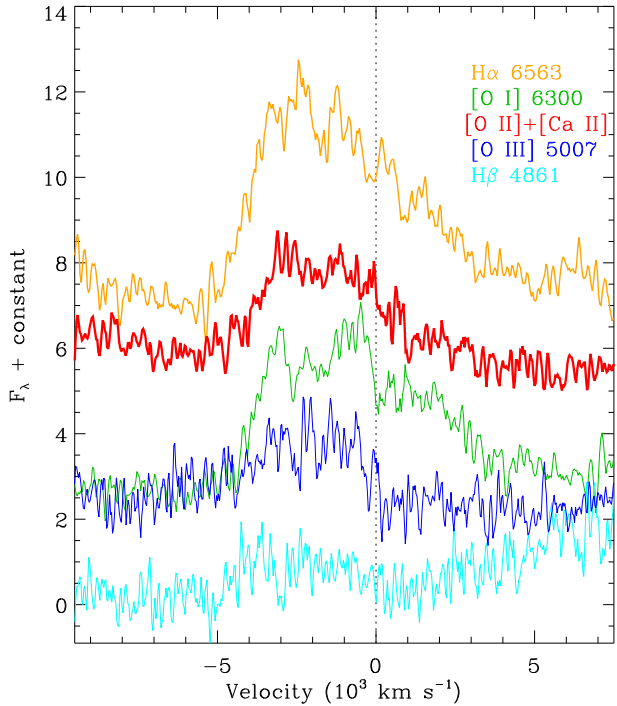}
\caption{Velocity profiles for the strongest lines in the spectrum of
  SN 1998S. The shortest wavelength transitions, particularly [O {\sc i}]
  $\lambda$6300 and [O {\sc iii}] $\lambda$5007, exhibit
  the most extreme flux cutoff on the red side of the line at zero
  velocity, strong evidence that internal dust extinction is the cause
  of the blue-shifted profiles. Note, the velocity of [O {\sc ii}]$+$[Ca {\sc ii}] should be taken with caution, since the center of the line blend could not be accurately determined.}
\label{fig:spec_vel}
\end{figure}
 
  \begin{figure}
\includegraphics[width=3.4in]{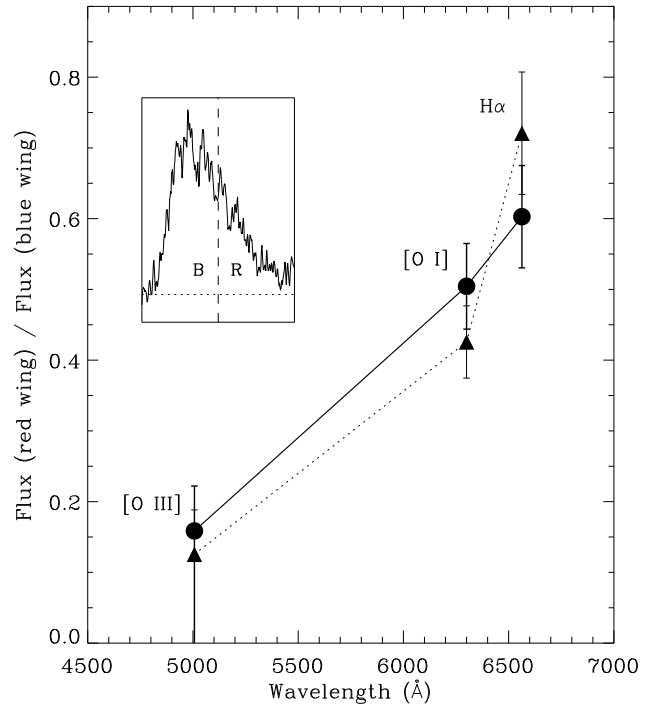}
\caption{Ratio of the red-wing to blue-wing flux plotted against wavelength for SN 1998S (\textit{circles}) and SN 1980K (\textit{triangles}). The [O {\sc iii}] value for SN 1998S could be taken as an upper limit, since its red wing appears indistinguishable from pure continuum. The inset illustrates the blue and red wing regions of the H$\alpha$ line of SN~1998S, centered on the rest wavelength.}
\label{fig:linerat}
\end{figure}
  
\subsection{Line Profiles and Dust Extinction}

The emission lines in the spectra SN~1998S exhibit
substantial asymmetry and a net blueshift of their centroids. A
commonly proposed explanation for this phenomenon is extinction by
dust. In the case of SNe that interact with dense CSM, dust is expected to form in the cool dense shell (CDS) that develops between the forward and reverse shocks. Emission lines formed in the post-shock gas, exterior to the dusty CDS, will suffer little or no extinction, while the same emission features emerging from the receding hemisphere will be obscured by the dust. This results in the suppression of the red wings of the emission lines, causing a net blueshift. Dust formation in the CDS has been postulated to explain the spectral evolution of the Type II SNe 1980K, 1979C, 1993J, and 1999em (Dwek et al. 1983; Fesen et al. 1999; Wang \& Hu 1994; Elmhamdi et al. 2003), the Type IIn SNe 2005ip and 2010jl (Smith et al. 2009b, 2012), and the Type~Ib/Ibn SN 2006jc (Smith et al. 2008). Blueshifted emission lines have also been observed to develop in the spectra of core-collapse SNe that do not exhibit evidence for immediate interaction with dense CSM. In fact, SN 1987A was the first case to provide a convincing connection between the development of emission-line asymmetries and dust formation. After about day 500, a drop in optical flux was accompanied by a rise in IR emission; while around the same time, emission lines from SN 1987A began to exhibit asymmetry and blueshift  (Danziger et al. 1989; Lucy et al. 1989; Colgan et al. 1994; Wang et al. 1996). Similar phenomena was observed in the photometric and spectral evolution of SN 1990I (Elmhamdi et al. 2004). In these cases, however, dust formation likely occurred within the metal rich ejecta.  

A large dust component has already been postulated in the case of SN
1998S. Gerardy et al. (2000) and Pozzo et al. (2004) monitored the NIR
photometry and spectra between days 305--1242, observing the onset of
CO emission and the development of a significant IR excess in the
spectral energy distribution. These features could be attributable to 
thermal emission from dust synthesized in the SN, or IR echoes reflecting off 
of pre-existing dust clouds in the outer CSM (see Sugerman et al. 2012), or both.  However, 
evolving asymmetry and increasing blueshift of the 
H$\alpha$ and He {\sc i} $\lambda$10830 lines observed by Pozzo et al. (2004)
provided evidence that dust is likely being synthesized within the SN CDS.

The emission-line profiles of SN 1998S provide strong evidence for internal dust.
Figure~\ref{fig:spec_vel} provides a close-up of the most prominent lines plotted in velocity coordinates. All of the emission lines exhibit significant blueshift and appear suppressed on their red wings. Interestingly, a relatively abrupt drop in flux at zero velocity is observed for the two shortest wavelength transitions of oxygen. The effect is most extreme for [O {\sc iii}] $\lambda$5007, which exhibits an apparently complete suppression of its red wing. Figure~\ref{fig:linerat} shows the ratio of red-wing to blue-wing flux for [O {\sc iii}], [O {\sc i}], and H$\alpha$ (H$\beta$ was excluded since [O {\sc iii}] overlaps with its red wing, and [O {\sc ii}]$+$[Ca {\sc ii}] was excluded because the central wavelength of the uncertain line blend could not be accurately determined). The trend could possibly be the result of wavelength dependent extinction. The flux drop between H$\alpha$ and [O {\sc iii}] could imply a relative extinction ratio of $\approx$1.4 mag between $\lambda$5007 and $\lambda$6563, which is consistent with an $R\approx3$ extinction coefficient for the obscuring material (Cardelli, Clayton, \& Mathis 1989). However, we stress caution in drawing any definitive quantitative conclusions about the extinction law from the line profiles. In fact, the relative blue- and red-wing flux ratios from one atomic transition to another could be affected by differences in the radial distributions of the [O~{\sc i}], [O~{\sc iii}], and H$\alpha$ emission zones. We discuss this further below. 

\subsubsection{Evidence for Dust in the Unshocked Ejecta}
The relative weakness of [O~{\sc iii}] emission, and the complete suppression of its red wing imply some interesting possibilities 
for the distribution of obscuring dust within SN 1998S. Since the [O {\sc iii}] emission is expected to be generated exclusively on the outer skin of the unshocked ejecta (CF94), interior to the CDS and reverse shock, then we expect dust formed in the CDS will obscure both sides of this emission line equally. Thus, the large net blueshift of the [O~{\sc iii}] emission line and the lack of a red wing require the obscuring material to be located interior to the [O~{\sc iii}] zone, within the \textit{unshocked} SN ejecta. If so, then practically the entire receding hemisphere of the [O~{\sc iii}] emission skin would be obscured by dust in both the inner ejecta \textit{and} in the CDS, explaining the lack of red wing flux. By comparison, the receding components of H$\alpha$ and [O {\sc i}] emission, both of which can be generated at larger radii in the post-shock gas, and also potentially echoed by the CSM, would suffer less complete obscuration from inner ejecta dust, resulting in more red-wing flux for these lines relative to [O~{\sc iii}]. Since both the blue and red wings of [O~{\sc iii}] are absorbed by dust within the CDS, this could explain why late-time [O~{\sc iii}] emission from SNe such as SN 1998S, 1980K, and 1979C, is often weaker than models predict (CF94; Fesen et al. 1999). 

The spectrum of SN 1998S thus provides compelling evidence for dust formation in the CDS and within the unshocked ejecta. Dust formation in the latter zone has also been postulated in the cases of SNe 2005ip and 2006jc (Smith et al. 2008, 2009b). In addition, mid-infrared observations of the SN remnant 1E 0102.2$-$7219 in the Small Magellanic Cloud (Rho et al. 2009; Sandstrom et al. 2009) have provided evidence for the formation of dust within unshocked SN ejecta, interior to the [O {\sc iii}] emission zones.   

\subsection{Evolution of Multi-peaked Oxygen Profiles}
\begin{figure}
\includegraphics[width=3.4in]{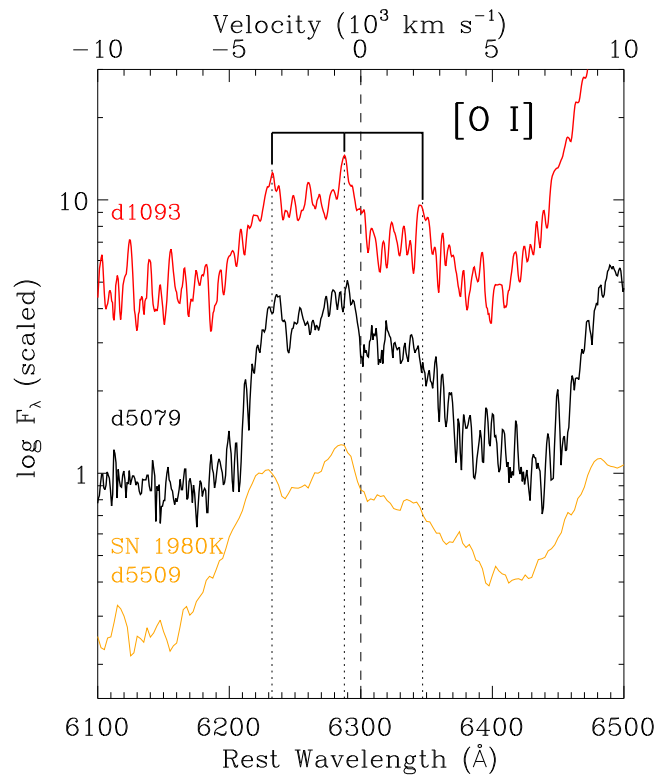}
\caption{Close-up of the [O {\sc i}] $\lambda$6300, 6364 profile on days
  1093 and 5079. The dashed line marks the rest wavelength of the
  $\lambda$6300 component of the transition. The dotted lines mark the center wavelengths/velocities of the
  triple peaks, assuming they all represent $\lambda$6300 (see caveats in text). SN 1980K on day 5509 exhibit very similar peak structure.}
\label{fig:ozoom}  
\end{figure}

The oxygen emission lines in the spectrum of SN 1998S exhibit
interesting higher-order structure.
Figure~\ref{fig:ozoom} shows a close-up of the [O {\sc i}] emission
feature on days 1093 and 5079. On day 1093, the line exhibits three
narrow peaks with maxima at $\lambda$6233, $\lambda$6287, and $\lambda$6346, all on top of a broader underlying profile. By day 5079,  the reddest of these peaks is no longer detected, leaving behind a double-peaked asymmetric profile having an overall blueshift, similar in appearance to SN 1980K on day 5509. As noted in \S3, although we cannot completely rule out the possibility that the reddest peak on day 1093 is a blueshifted $\lambda$6364 component of the doublet, we suspect that it actually represents a redshifted $\lambda$6300 component.  Assuming this to be the case, the blue and red peaks would have velocities of approximately $\pm$2600--2700
km s$^{-1}$ with respect to the central peak. The origin of these peaks 
and the cause of their temporal change is puzzling. 

Double-peaked oxygen emission features have been observed in many
other core-collapse SNe (e.g., see Modjaz et al. 2008, and references therein; Milisavljevic et al. 2009). One 
interpretation involves emission from large individual clumps of 
oxygen-rich ejecta (Wang \& Chevalier 2002). However, this model 
could have difficulty explaining the apparent similarity of the wavelength 
spacing between peaks that is typically seen from one SN to another. Other interpretations include emission from the 
equatorial ring or torus of a bipolar outflow viewed nearly edge-on 
(Chugai \& Danziger 1994; Maeda et al. 2008). However, the toroidal model always produces 
two peaks centered around the rest velocity, in addition to a possible central component.  Milisavljevic et al. (2010) pointed out that the double-peaked oxygen profiles observed in core-collapse SNe (mainly Type Ib, IIb, and Ic) 
typically come in two main groups: \textit{symmetric} profiles, where the trough between
two peaks lies near rest velocity, with the peak separation lying conspicuously close the the 64 {\AA} separation of the doublet components; and \textit{asymmetric} profiles,
where one component lies near rest and the other is blueshifted,
having no conspicuous redshifted counterpart. The [O {\sc i}] profile
of SN~1998S on day 5079 falls into the latter asymmetric category. 
On day 1093, however, the line appears more symmetric with the inclusion of the conspicuous red peak.  
The subsequent disappearance of this peak by day 5079 implies that it has been extinguished by dust, 
and therefore, originates from the receding hemisphere of the SN. This supports the interpretation that the peak is 
a redshifted $\lambda$6300 emission feature from the doublet, and not a blueshifted $\lambda$6364 feature. 
 
We question whether the [O {\sc i}] profile evolution could be related to the emergence and disappearance of the triple-peaked H$\alpha$ profile at earlier times. By day 258, the H$\alpha$ profile of SN~1998S exhibited a dramatic
triple-peaked profile (see Figure~\ref{fig:spec_comp}). Its red and blue peaks were shifted by roughly
$\pm$4000 km s$^{-1}$ with respect to the central peak. Spectra from subsequent epochs exhibited a continuous flux decrease
of the red peak, attributed to increasing extinction of the receding hemisphere of the SN by dust. By day $\approx$650 the red
peak had become extinguished below detectability. The evolving red and blue peaks have been interpreted as the result of
emission from an expanding ring or torus, with the central peak possibly arising from shocked circumstellar clumps (Gerardy et al. 2000; Leonard et al. 2000). However, Fransson et al. (2005) identified inconsistencies between this model and the observations, and instead favored a scenario in which double-peaked emission profiles emerge from a geometrically thin CDS behind the reverse shock, at large optical depth. Still, the CDS would probably have to be non-spherical to be consistent with the polarization measurements by Leonard et al. (2000). In any case, the observed evolution of the triple-peaked H$\alpha$ profile motivates the following question.  Could the triple-peaked structure of [O~{\sc i}] $\lambda$6300 seen on day 1093 --- and the subsequent disappearance of its red component by day 5079 --- be attributable to similar processes involving ring-like/toroidal ejecta or a non-spherical thin CDS, where the red peak of a corresponding triple-peaked profile eventually becomes extinguished as a result of persistent dust formation? Indeed, whatever structure the reverse-shock/CSM interaction exhibited at relatively early times would likely manifest again later as oxygen-rich ejecta eventually crosses over.

Disk, ring, and toroidal structures in SN ejecta have been suggested as viable explanations for observations of oxygen-rich SN remnants such as SNR 1E 0102.2$-$7219 in the Small Magellanic Cloud and Cassiopeia A in the Galaxy, both of which exhibit velocity structures indicative of asymmetric explosion geometries (Tuoho \& Dopita 1983); indeed, the integrated spectra of the [O {\sc i}]  $\lambda$6300 line in Cassiopeia A exhibits a multi-peaked structure similar to what is seen in the spectra  SN 1998S and 1980K (Milisavljevic et al. 2012).  In addition, spectropolarimetry of other core-collapse SNe (Wang \& Wheeler 2008, and references therein), the morphologies 
 of many young Galactic remnants, and pulsar space velocity distributions (Cordes, Romani, \& Lundgren 1993; Lyne \& Lorimer 1994; Cordes \& Chernoff 1998), collectively support the commonality of asymmetric explosions from core-collapse SNe.   \medskip
 
\section{conclusions}
Our observations of the IIn SN 1998S show that the CSM interaction is still going strong 14 years after explosion.
The H$\alpha$ luminosity has hit a slowly declining floor of $L=3.1\times10^{37}$ erg s$^{-1}$ ($\approx$8000 $L_{\odot}$). 
This behavior is very similar to SN 1980K, but indicates that the pre-SN mass-loss rate for the RSG progenitor of SN 1998S was  extraordinarily high, even $\approx$1200 years before core collapse. This behavior suggests that other extreme RSGs, such as VY CMa, can potentially endure a similarly high degree of mass loss for at least 1200 years before exploding. 

The blueshifted emission-line profiles observed from SN 1998S provide compelling evidence for the formation of dust. The relatively strong extinction of the red wing of the [O~{\sc iii}] $\lambda$5007 line, in particular, is very interesting, as this is either a consequence of the wavelength dependence of the extinction, or the fact that the [O~{\sc iii}] emission is generated exclusively interior to the shocks on the ionized skin of the inner ejecta, having its receding hemisphere completely obscured by inner core dust. Evidence for dust within the unshocked ejecta of SNe, interior to the [O~{\sc iii}] emission zone, was acquired from mid-infrared observations of the SN remnant 1E 0102.2$-$7219 in the Small Magellanic Cloud (Rho et al. 2009; Sandstrom et al. 2009). Dusty O-rich remnants such as this could thus provide a glimpse at the possible appearance of SN 1998S and related SNe, centuries into the future.  

Finally, the observed evolution of the multi-peaked [O~{\sc i}] profile, particularly the disappearance of the individual red peak, suggests that the red peak originated within the receding hemisphere of the SN and became obscured by dust that formed between days 1093 and 5079. The peak would therefore have to represent the $\lambda$6300 component of the doublet. This gives some credence to the interpretation that the individual peaks of oxygen profiles observed in other SNe represent emission features from separate structures within the outflow. This also implies that the ``missing" conspicuous red peak in the multi-peaked [O {\sc i}] profiles from other SNe might be the result of varying degrees of dust obscuration. Interestingly, SN 1998S must have formed significant amounts of dust after 3 years in order to extinguish the red peak between days 1093 and 5079.

\section*{acknowledgements}

This research was based in part on observations made with the
Large Binocular Telescope. The LBT is an international collaboration
among institutions in the United States, Italy and Germany. The LBT
Corporation partners are: the University of Arizona on behalf of the
Arizona university system; the Istituto Nazionale di Astrofisica,
Italy; the LBT Beteiligungsgesellschaft, Germany, representing the Max
Planck Society, the Astrophysical Institute Potsdam, and Heidelberg
University; the Ohio State University; and the Research Corporation,
on behalf of the University of Notre Dame, University of Minnesota and
University of Virginia. Our analysis utilized archival spectra of
 SN~1998S from the SUSPECT database. We thank T. Matheson, R. Fesen and D. Milisavljevic
for providing earlier spectra of SN 1998S and 1980K for our comparison. We thank the
referee Nikolai Chugai for insightful comments, which helped to improve the manuscript.

\end{document}